
\input phyzzx
\hfuzz 25pt

\def\dplus{=\hskip-5pt \raise 0.7pt\hbox{${}_\vert$} ^{\phantom 7}}
\def\dplusup{=\hskip-5.1pt \raise 5.4pt\hbox{${}_\vert$} ^{\phantom 7}}
\def\dplus{=\hskip-4.8pt \raise 0.7pt\hbox{${}_\vert$} ^{\phantom 7}}

\def\pmb#1{\setbox0=\hbox{#1} \kern-.025em\copy0\kern-\wd0
\kern0.05em\copy0\kern-\wd0 \kern-.025em\raise.0433em\box0}

\font\mybb=msbm10 at 10pt

\def\bb#1{\hbox{\mybb#1}}

\def\bE{\bb {E}}

\REF\strom{A. Strominger, {\sl Open p-branes}, Phys. Lett. 
{\bf B383} 44; hep-th/9512059.}
\REF\town{P.K. Townsend, {\sl Brane Surgery}, Nucl. 
Phys. Proc. Suppl {\bf 58} (1997) 163.}
\REF\pt{G. Papadopoulos and P.K. Townsend, 
{\sl Intersecting M-branes}, Phys. Lett. {\bf B380}
(1996) 273.}
\REF\callan {C.G. Callan, Jr and J.M. Maldacena, 
{\sl Brane dynamics from the
Born-Infeld action}, hep-th/9708147. }
\REF\gibbons{G.W. Gibbons, {\sl Born-Infeld Particles 
and Dirichlet p-branes}, hep-th/9709027.}
\REF\howeb{P.S. Howe, N.D. Lambert and P. West, 
{\sl The self-dual string soliton},
hep-th/9709014.} 
\REF\howe{P.S. Howe, N.D. Lambert and P. West, 
{\sl The  three-brane soliton of the M-five-brane},
hep-th/9710033; {\sl Classical M-five-brane dynamics 
and quantum N=2 Yang-Mills}, hep-th/9710034.}
\REF\seiberg{N. Seiberg {\sl A Matrix Description 
of M-theory on $T^5$ and $T^5/Z_2$},
hep-th/9705221.}
\REF\seibergb{N. Seiberg and S. Sethi, {\sl Comments 
on Neneu-Schwarz fivebranes}, hep-th/9708085.}
\REF\brunner{I. Brunner and A. Karch, {\sl Matrix
 Description on M-theory on $T^6$},
hep-th/9707259.}
\REF\wittenb{E. Witten, {\sl New \lq\lq gauge theories" 
in six-dimensions}, hep-th/9710065.}
\REF\nil{M. Cederwall, A. von Gussich, B.E.W. 
Nilsson, P. Sundell and A. Westerberg,
{\sl The Dirichlet Super P-Branes in Ten-Dimensional
 Type IIA and IIB Supergravity}, Nucl. Phys.
{\bf B490} (1997) 179.}
\REF\aga{M. Aganagic, C. Popescu and J.H. Schwarz, 
{\sl D-Brane Actions with Local Kappa
Symmetry}, Phys. Lett. {\bf B393} 311.}
\REF\bergb{E. Bergshoeff and P.K. Townsend, 
{\sl Super D-branes}, Nucl. Phys. {\bf B490} (1997)
145.}
\REF\howec{P.S. Howe and E. Sezgin, {\sl D=11, p=5}, 
Phys. Lett. {\bf B390} (1997) 62.}
\REF\pasti{I. Bandos, K. Lechner, A. Nurmagambetov, 
P. Pasti, D. Sorokin and M. Tonin, {\sl
Covariant Action for the Super-Five-Brane of M-theory}, 
hep-th/9701149.}
\REF\agab{M. Aganagic, J. Park, C. Popescu and J.H. 
Schwarz, {\sl Worldvolume Actions of
the M-theory Five-Brane}, hep-th/9701166.}
\REF\witten{E.Witten {\sl Solutions of four-dimensional 
field theories via M-theory}, Nucl. Phys.
{\bf B500} (1997) 3; hep-th/9703166.}
\REF\erictwo{E. Bergshoeff, B. Janssen and T. 
Ort\'in, {\sl Kaluza-Klein Monopoles
and Gauged Sigma-Models}, hep-th/9706117.}
\REF\papac{E. Bergshoeff, R. Kallosh, 
T. Ort\'in and G. Papadopoulos, {\sl $\kappa$-Symmetry,
Supersymmetry and Intersecting Branes}, hep-th/9705040.}
\REF\papad{G. Papadopoulos and P.K. Townsend, 
{\sl Kaluza-Klein on the Brane}, 
Phys. Lett. {\bf
B393} (1997) 59.}
\REF\lambert{J.M. Izquierdo, N.D.
 Lambert, G. Papadopoulos and P.K. Townsend, {\sl Dyonic
Membranes}, Nucl. Phys. {\bf B460} (1996) 560, hep-th/9508177.}
\REF\schwarz{J.H. Schwarz,{\sl An $SL(2,Z)$ 
multiplet of type IIB superstrings}, Phys. Lett. {\bf
B360} (1995) 13 (E: {\bf B364} (1995) 252), 
hep-th/9508143, hep-th/9509148.}
\REF\tseytlinb{J.G. Russo and A.A. Tseytlin, 
{\sl Waves, boosted branes and BPS states in
M-theory}, Nucl. Phys. {\bf B490} (1997) 121,
 hep-th/9611047.}
\REF\costa{M.S. Costa and G. Papadopoulos, 
{\sl Superstring dualities and p-brane bound states},
 hep-th/9612204.}
\REF\ohta{N. Ohta and J-G Zhou, {\sl Towards 
the classification of non-marginal bound states of
M-branes and their construction rules}, hep-th/9706153.}
\REF\sorkin{R. Sorkin, {\sl Kaluza-Klein
 monopole}, Phys. Rev. Lett. {\bf 51} (1983) 87.}
\REF\gperry{ D.J. Gross and M. Perry, {\sl
 Magnetic monopoles in Kaluza-Klein theory}, Nucl.
Phys. {\bf B226} (1983) 29.} 
\REF\hawkinga{S.W. Hawking, {\sl Gravitational 
Instantons}, Phys. Lett. {\bf 60A} (1977) 81.} 
\REF\hawkingb{G.W. Gibbons and S.W. Hawking, 
{\sl Gravitational Multi-Instantons}, Phys. Lett. {\bf
B78} (1978) 430.}
\REF\townb{E. Bergshoeff, J. Gomis and P.K. Townsend, 
{\sl M-brane Intersections from worldvolume
superalgebras}, hep-th/9711043.}

\Pubnum{ \vbox{ \hbox{R/97/64}\hbox{} } }
\pubtype{Revised, June 1998}
\date{December, 1997}
\titlepage
\title{T-duality and the Worldvolume Solitons of Five-Branes and KK-Monopoles}
\author{G. Papadopoulos}
\address{\negthinspace DAMTP \break Silver Street \break Cambridge CB3 9EW}
\abstract {We show that the fluxes
of the various six-dimensional \lq\lq gauge" theories are 
associated to below threshold bound states
 of D-branes with the  NS-5-branes   and  
KK-monopoles which 
preserve half of bulk supersymmetry. We 
then present the supergravity solutions
that correspond to these bound states. In addition using the
 worldvolume solitons of
IIA and IIB NS-5-branes   and  KK-monopoles, 
we investigate the sectors of the \lq\lq gauge"
theories that preserve one quarter of bulk supersymmetry. This 
leads to a generalization a supergravity solution
which   has the interpretation of two intersecting 
NS-5-branes at a 3-brane and to the
construction of some of the worldvolume solitons of  
IIA and M-theory KK-monopoles.  Furthermore,
using the IIA/IIB T-duality of  the bulk theories,
 we give the T-duality transformations of the
worldvolume solitons of NS-5-branes and KK-monopoles. 
We find that  the worldvolume 0-brane,
self-dual string and 2-brane solitons of  NS-5-branes
 appear in the same T-duality chain.  }

\endpage
\pagenumber=2
\font\mybb=msbm10 at 12pt
\def\bb#1{\hbox{\mybb#1}}

\def\C{\mkern1mu\raise2.2pt\hbox{$\scriptscriptstyle|$}\mkern-7mu{\rm C}}

\def\log{{\rm log}}

\sequentialequations

\chapter{Introduction}

It has been known for sometime that the
 boundary of a brane ending on another
brane [\strom, \town] and intersection 
region of two or more branes
are described by worldvolume 
solitons [\pt].   These
solitons usually preserve a subgroup 
of the Poincar\'e group of the 
branes involved and therefore they have 
themselves a brane interpretation. 
 Many such worldvolume
solitons have been found  in [\callan, 
\gibbons, \howeb] and their supersymmetry
has been investigated in [\howe].

More recently there is some interest in  
six-dimensional \lq\lq gauge"
theories which are found as limits of the 
superstring theory in the presence of $n$ 
IIB NS-5-branes as the
string coupling constant goes to zero. In this limit the
 theory at the  bulk becomes free IIB
string theory but interactions persist on the  NS-5-branes 
leading to an interacting six-dimensional \lq\lq gauge"
theory [\seiberg, \seibergb, \brunner].   The effective 
theory is described by a (1,1) supersymmetric 
Yang-Mills multiplet with gauge group $SU(n)$
(factoring out the centre of mass)
which  is a reduction of the
ten-dimensional super-Yang-Mills multiplet to six dimensions.
Some of the excitations of the six-dimensional \lq\lq gauge" theory  are due to
D-branes that lie within the NS-5-branes. 
All these excitations but one
of this six-dimensional \lq\lq gauge" 
theory can be identified with 
fluxes of the effective theory\foot{In the Coulomb branch
of this theory the gauge group 
is broken to a product of $U(1)$'s.}; the
remaining degree of freedom signals that the
 six-dimensional theory has another sector which
is not captured by the effective theory.

There are two possible IIA (or M-theory) duals to the
 above six-dimensional \lq\lq gauge" theory associated with
the IIB strings.
\item{(i)} Using T-duality along one of
worldvolume directions of the IIB NS-5-brane leads to a IIA 
six-dimensional theory associated with the 
IIA NS-5-brane. The effective theory 
is described  by a
$SU(n)$ generalization of (2,0)-supersymmetric
 tensor multiplet\foot{An action for the  non-abelian 
(2,0)-supersymmetric multiplet is not known.} in six 
dimensions.  Some of degrees of
freedom  of this theory are again due to IIA
 D-branes that lie within the IIA
 NS-5-brane. All these degrees of freedom can be identified with the
fluxes of the 3-form self-dual field strength and 
the fluxes of a 5-form field strength; the latter is
dual to the scalar along the compact eleventh
 direction\foot{The IIA NS-5-brane is viewed as the
reduction of the  M-5-brane along a transverse direction.}.
\item{(ii)} Using T-duality 
along a  transverse direction of the IIB NS-5-brane 
leads to another six-dimensional \lq\lq gauge" theory
 associated with the IIA Kaluza-Klein (KK) monopole.  In
this case the effective theory is described by the  six-dimensional
  (1,1) supersymmetric
Yang-Mills multiplet
 which is the reduction of the N=1 ten-dimensional
one with gauge group $SU(n)$ on a four-torus.

A larger class of six-dimensional theories can be found by
 starting from the (p,q)-5-branes
of IIB and then by T-dualizing in a transverse direction  
followed by a lifting to M-theory. The
effective  theory is described by a
(1,1)-supersymmetric Yang-Mills multiplet
with gauge group $SU(r)$ where $r$ is the
 largest common divisor of $p$ and $q$ [\wittenb].

In this letter, we shall show that all the configurations that
 involve D-branes that lie within
NS-5-branes are bound states below threshold which preserve $1/2$ of 
bulk supersymmetry. Consequently, the
excitations of the  six-dimensional \lq\lq gauge" theories
  due to these bound
states also preserve $1/2$ of supersymmetry. 
 We shall verify that all the required
 bound states exist as solutions of 
IIA and IIB supergravity theories preserving
 $1/2$ of  supersymmetry.  
We shall also present the worldvolume analogue of
 these bound states. In particular, we shall find
 solutions  of the field equations of Born-Infeld
(BI) actions\foot{The dynamics of a IIB NS-5-brane 
is described by a BI action as it is related to
D-5-brane by S-duality. The field equations of M-theory 
five-brane, instead of the usual two-form BI
field strength, involve a 3-form self-dual 
field strength.} [\nil-\agab] of IIB  and IIA
NS-5-branes which preserve
$1/2$ of bulk supersymmetry and have the  fluxes 
as described  in [\seiberg].

The BI actions,  apart from the
solutions that preserve
$1/2$ of bulk supersymmetry,  
admit solitons which preserve $1/4$ or less of
supersymmetry. 
 In the context of six-dimensional \lq\lq gauge" 
theories, these
 solitons describe  sectors 
which preserve $1/2$, $1/4$ or less of
supersymmetry.
We shall refer to these  sectors as $1/2$-sector, 
$1/4$-sector and so on.  We shall argue using supersymmetry
that some of the worldvolume solitons cannot propagate in the bulk.
We shall then explore the relation
between worldvolume solitons and intersecting branes and compare the
two pictures. As a result we shall find a  new supergravity
solution which has the interpretation of two IIA or IIB NS-5-branes
intersecting on a 3-brane and corresponds to the 3-brane soliton 
of [\howe]. 

We shall also investigate the six-dimensional
 \lq\lq gauge" theories based
on the IIA and IIB KK-monopoles.  As in the case 
of NS-5-branes, the $1/2$-sector 
is associated  bound states of D-branes with the 
IIA and IIB KK-monopoles which are below
threshold and preserve $1/2$ of bulk supersymmetry.
 We shall also use T-duality to describe
the associated supergravity solutions.  Then 
 using the  worldvolume action of the M-theory
KK-monopole [\erictwo], we shall
 find some of worldvolume brane solitons of the 
IIA and M-theory  KK-monopoles which
preserve $1/4$ of bulk supersymmetry.
In addition,  viewing the worldvolume 
solitons of NS-5-branes and KK-monopoles 
as brane boundaries and intersections, 
we shall use the type II T-duality rules 
of the bulk theories
to find the T-duality transformations of 
the worldvolume solitons. It turns out 
that worldvolume solitons that appear in 
the same T-duality chain have similar solutions.
An example of this is  the 
0-brane, the self-dual string and the 2-brane 
worldvolume solitons of the IIA and
IIB  NS-5-branes.

This letter is organized as follows. In sections
 two and three, we investigate the $1/2$-and
$1/4$-sectors of the six-dimensional theories 
associated with the NS-5-branes, respectively. In
sections four and five, we examine the $1/2$-and
$1/4$-sectors of the six-dimensional theories 
associated with the KK-monopoles, respectively. In
section six, we give the T-duality transformations 
of the worldvolume solitons and finally in
section seven we present our conclusions.


\chapter{D-branes within the IIA and IIB NS-5-branes }

\section{Supergravity}

The bound states that involve D-branes 
and NS-5-branes in IIA theory are
$(0_D|5_S)_A$, $(2_D|5_S)_A$ and $(4_D|5_S)_A$\foot{In this notation, 
$(0_D|5_S)_A$ denotes the
bound state of a D-0-brane with a solitonic (NS) 5-brane in 
IIA theory and similarly for the rest
of the bound states.}. 
Similarly, the IIB D-brane/NS-5-brane bound  states
 are $(1_D|5_S)_B$, $(3_D|5_S)_B$ and $(5_D|5_S)_B$.  All these 
bound states preserve $1/2$  of
bulk supersymmetry provided that the associated D-brane is within
 the NS-5-branes and are below
threshold\foot{ We remark that  if the D-branes are placed parallel 
but not within the NS-5-brane, then all supersymmetry is broken.}.
This can be easily seen by, for example, examining the 
supersymmetry projectors associated with the
branes involved in the bound states (see  [\papac,
\papad]).  The energy of these bound states [\lambert] is 
consistent with the formula given in [\seiberg] for
computing the energy of the excitations of the six-dimensional
\lq\lq gauge" theories from D-branes within NS-5-branes.

All the supergravity solutions that correspond to the
above bound states can be found using the T-duality chain
$$
(0_D|5_S)_A{\buildrel  T\over\leftrightarrow} 
(1_D|5_S)_B {\buildrel  T\over\leftrightarrow}
(2_D|5_S)_A
{\buildrel  T\over\leftrightarrow} 
(3_D|5_S)_B{\buildrel  T\over\leftrightarrow} (4_D|5_S)_A
{\buildrel  T\over\leftrightarrow} (5_D|5_S)_B\ ,
\eqn\infivee
$$
where  T-duality operations are performed along  NS-5-brane worldvolume directions. 
These solutions
are controlled by one harmonic function which 
indicates that the positions of the branes
involved in the superposition coincide.
Some of these solutions have already being found, for 
example the $(2_D|5_S)_A$ has been given in [\lambert]
and the $(5_D|5_S)_B$ has been given in [\schwarz].  The $(4_D|5_S)_A$ 
is most easily
constructed by reducing the M-theory five-brane 
at an angle (see [\tseytlinb-\ohta]). The remaining solutions 
 can be  constructed from the T-duality chain
and they will not be presented here.

\section{Worldvolume Solutions}

The effective theory of the IIB NS-5-brane is described by a BI action
which in the static gauge is
$$
I= \int\, d^6u\, \sqrt{|{\rm det}
\big(\eta_{\mu\nu}+\partial_\mu X^i \partial_\mu
X^j \delta_{ij}+F_{\mu\nu}\big)|}\ ,
\eqn\insevenn
$$
 where $\{X^i; i=1, \dots, 4\}$ are the 
transverse coordinates of the 5-brane, $\{u^\mu;
\mu=0,\dots,5\}$ are the worldvolume 
coordinates and $F_{\mu\nu}$ is the BI field. The worldvolume
 solutions that corresponds to the  bulk bound states
\infivee\ is
$$
\eqalign{
F_{\mu\nu}&=a^I \alpha_I
\cr
X^i&=c^i\ ,}
\eqn\constconf
$$ 
where $\{\alpha_I;I=1, \dots, 15\}$ is a basis of two 
forms on $\bE^{(1,5)}$, and $\{a^I; I=1,
\dots, 15\}$ and $\{c^i; i=1, \dots, 4\}$ are 
real constants. It is a simple
 application of the results of [\papac] to show
that such configurations preserve $1/2$ of
 the supersymmetry of the bulk. The solution
 \constconf\ gives the fifteen fluxes\foot{Upon
compactification on a five-torus the fluxes are
 quantized.} in the counting of [\seiberg]. The
missing flux  corresponds to the D-5-brane within 
the NS-5-brane bound state. From the supergravity
point of view it is easy to see why such a flux
 cannot be found within  the effective theory. The
supergravity solution is invariant under the 
six-dimensional Poincar\' e group of the worldvolume of
the five branes.  So the corresponding solution of
 the effective theory should have the same degree
of symmetry but there are not exist such solutions 
of the effective theory apart from the
vacuum one.

 The field equations of the effective theory of 
IIA NS-5-brane can be found by reducing along a 
transverse direction those of M-5-brane 
[\howec-\agab]. We shall be interested in two 
classes of solutions which preserve $1/2$ of bulk supersymmetry. 
In the first case, we take the
BI field to vanish but allow the transverse scalar
 along the eleventh direction to depend linearly
on the worldvolume coordinates, {\sl i.e}
$$
\eqalign{
f_3&=0 
\cr
X^{11}&= b_\mu u^\mu+a
\cr
X^i&=c^i\ .}
\eqn\ineightt
$$
Such solution preserves $1/2$ of bulk supersymmetry and gives six  
fluxes as in
[\seiberg]. For the other solution,  we take all the 
transverse scalars to be constant and the BI
field to be constant and self-dual, {\sl i.e.}
$$
\eqalign{
f_3&=a^I \omega_I
\cr
X^i&=c^i\ ,}
\eqn\intenn
$$
where $\{\omega_I; I=1,\dots, 10\}$ is a basis of 
constant self-dual 3-forms on $\bE^{(1,5)}$. 
Such solution also preserves $1/2$ of bulk
 supersymmetry and  gives the remaining ten
fluxes of [\seiberg].


\chapter{The $1/4$-sector and Brane Solitons}

\section{Worldvolume Solutions}
 
To investigate the $1/4$-sector of
\lq\lq gauge" theories, we shall present the
worldvolume solutions of NS-5-branes that preserve $1/4$
of bulk supersymmetry. In the case of IIB NS-5-brane
these worldvolume solutions are
the 0-brane, the 2-brane and the 3-brane. 
Let $F_2$ and $X^1$ be the BI field
 and one of the transverse scalars of the IIB
NS-5-brane, respectively. The 0-brane and 
2-brane worldvolume solutions are
$$
\eqalign{
F_2&=-{1\over4} dt\wedge dH_{(5)}
\cr
X^1&=H_{(5)}\ ,}
\eqn\zerobrane
$$ 
and
$$
\eqalign{
F_2&=-{1\over4}\star dH_{(3)}
\cr
X^1&=H_{(3)}\ ,}
\eqn\twobrane
$$ 
respectively, where $H_{(n)}$ is a harmonic function on $\bE^n$ 
 and the Poincar\'e star operation
is with respect to the flat metric on the 
three-dimensional subspace of the NS-5-brane transverse
to the 2-brane soliton.
For the 3-brane worldvolume soliton, the BI 
field vanishes. If we set $z=u^4+iu^5$ and
$s=X^1+iX^2$, the 3-brane solution is determined 
by any holomorphic function $s=s(z)$ or in an
implicit form
$$
Z(s,z)=0\ .
\eqn\threebrane
$$
These solitons are the same as those found 
in the worldvolume theory
 of the D-5-brane [\callan,\gibbons,\howe]. In fact
they are related to them by S-duality.

The above
worldvolume brane solitons do not propagate 
in the bulk. This is 
straightforward for the case of
 0-brane and 2-brane solitons since there 
are no 0-branes and
 2-branes in IIB theory. To see this
for the case of the  3-brane soliton, we 
shall first argue that 
the 3-brane soliton cannot be
identified with the IIB D-3-brane. To show this, 
let us suppose that 
the 3-brane soliton of the
NS-5-brane is identified with the IIB 3-brane. 
If this were the case,
 we  consider the configuration
of a  D-3-brane parallel to the NS-5-brane 
separated with a distance
$r$. As we have seen in the previous 
section, for $r\not=0$ this
 configuration breaks  all
supersymmetry. For $r=0$, the  D-3-brane 
is within the NS-5-brane and
 $1/2$ of  bulk supersymmetry is preserved. 
Therefore a  3-brane 
worldvolume soliton of NS-5-brane
and IIB D-3-brane within the NS-5-brane bound state 
preserve  different fractions of bulk 
supersymmetry leading to
a contradiction of the original hypothesis. So the 
3-brane worldvolume soliton of NS-5-brane and the 
D-3-brane should no be identified. Since 
it is unlikely
to exist another supersymmetric D-3-brane 
in IIB theory, because there 
is a unique 5-form field
strength, the 3-brane worldvolume soliton
 does not propagate in the bulk.

Let $f_3$  be the self-dual 3-form field strength and $X^1$ be
one of the transverse scalars of worldvolume theory of
 IIA NS-5-brane.
There are two NS-5-brane worldvolume 
solitons preserving $1/4$ of bulk supersymmetry, the 
 self-dual string  [\howe]
$$
\eqalign{
f_3&={1\over4}\big[dt\wedge d\rho\wedge dH_{(4)}+\star dH_{(4)}\big]
\cr
X^1&=H_{(4)}}
\eqn\onebrane
$$ 
and  the 3-brane  whose
 solution is given as in \threebrane,
where $\{t,\rho\}$ are the  coordinates of 
the string and the Poincar\' e star
operation is with respect to the flat metric 
on the four-dimensional subspace of the
NS-5-brane transverse to the string.
Since there is no 3-brane in IIA theory, the 3-brane 
worldvolume soliton cannot propagate in the bulk. We have not
 being able to
find a similar argument for the self-dual string. 
It is worth mentioning that the expressions for
the solutions of the 0-brane, 2-brane and self-dual
 string worldvolume solitons are remarkably
similar. As we shall show in section six, 
they are related by T-duality.

\section{Supergravity}

As we have already mentioned in the introduction, 
from the bulk perspective, 
the various brane
solitons of the  effective theories can be
 realized as brane boundaries or brane
 intersections.
In the IIB case, the 0-brane, 2-brane and the 
3-brane worldvolume solitons on the NS-5-brane are
realized as follows:
The 0-brane soliton is the boundary of a D-string 
on a NS-5-brane, ($(0|1_D,5_S)_B$ in the notation
of [\papad]),  the 2-brane soliton is the
boundary of a IIB 3-brane on a NS-5-brane $(2|3_D,5_S)_B$, and 
the 3-brane soliton is the intersection of 
two NS-5-branes on a 3-brane $(3|5_S,5_S)_B$. 
The associated
supergravity configurations\foot{There is no supergravity
 solution that describes a brane ending on another one.
However there are some 
 solutions that  resemble such configurations.} are derived from
M-theory by the following chains of reductions and T-dualities:
$$
\eqalign{
(1|2,5)_M&{\buildrel  R\over\rightarrow} 
(1|2_D,5_S)_A{\buildrel  T\over\leftrightarrow}
(0|1_D,5_S)_B 
\cr
(1|2,5)_M&{\buildrel  R\over\rightarrow}
(1|2_D,5_S)_A{\buildrel  T\over\leftrightarrow}
(2|3_D,5_S)_B 
\cr
(3|5,5)_M&{\buildrel  R\over\rightarrow} 
(3|5_S,5_S)_A{\buildrel  T\over\leftrightarrow}
(3|5_S,5_S)_B\ . }
\eqn\tduality
$$

In the IIA case, the self-dual string is the boundary of a D-2-brane
ending on a NS-5-brane and the
3-brane soliton is  the intersection of two NS-5-branes. 
The associated supergravity
configurations can be constructed using
 the second and third chains above.

To illustrate further the relation between 
brane solitons and intersecting branes,
we shall present the (IIA and IIB) 
supergravity solution with the
interpretation of two intersecting NS-5-branes 
at a 3-brane which corresponds to the
soliton \threebrane.  For this we
shall take the overall positions of the two
 NS-5-branes to coincide. Let $F_3$ and $\phi$ be the
3-form field strength of the NS-NS sector 
and the dilaton, respectively. This solution can be
written as
$$
\eqalign{
ds^2&= ds^2\big(\bE^{(1,3)}\big)+H_1 
ds^2(\bE^2)+H_2 ds^2(\bE^2)+ H_1 H_2 ds^2(\bE^2)
\cr
F_3&= \omega_1(\bE^2)\wedge {}^*dH_1+\omega_2(\bE^2)\wedge {}^*dH_2
\cr
e^{2\phi}&=H_1 H_2\ ,}
\eqn\twonee
$$
where $\omega_1$ and $\omega_2$ are the
 volume forms of the relative transverse
directions, the Poincar\'e star operation is 
with respect to the flat metric of the overall
transverse directions, and 
$H_1$ and
$H_2$ are the real and imaginary parts of a  
holomorphic function of the overall
transverse coordinates $(z, \bar z)$, 
{\sl i.e.} $Z(s,z)=0$ where $s=(H_1+iH_2)(z)$. 
In particular, if we set $t=e^{-s}$. 
 The simplest
choice is $s=-\sum_i q_i \log(z-z_i)$ or 
equivalently the zero locus of the polynomial
$F(t,z)= t^2- \Pi_i (z-z_i)^{2q_i}$.
A more general configuration is given by the zero locus of the polynomial
$$
F(t,z)=t^2-2B(z) t+\Lambda^{2N}
\eqn\wittensol
$$
with scale $\Lambda$ used in [\witten], where
$$
B(z)=z^N+u_{N-1} z^{N-2}+u_{N-2} z^{N-3}+\dots +u_1\ ,
\eqn\twofourr
$$
and  $\{u_i; i=1,\dots, N-1\}$ are the 
moduli parameters of a Riemann surface.

\chapter{Branes within the IIA and IIB KK monopoles }

 From the bulk perspective, the  IIA D-branes within 
the IIA KK-monopole bound states
 which preserve
$1/2$ of  supersymmetry can be found by acting with
 T-duality on the D-brane/NS-5-brane bound
states of IIB theory along a direction transverse to the NS-5-brane
as 
$$
\eqalign{
(1_D|5_S)_B&{\buildrel  T\over\leftrightarrow }(2_D|0_M)_A
\cr
(3_D|5_S)_B&{\buildrel  T\over\leftrightarrow} (4_D|0_M)_A
\cr
(5_D|5_S)_B&{\buildrel  T\over\leftrightarrow} (6_D|0_M)_A\ ,}
\eqn\twofivee
$$
where $0_M$ denotes the KK-monopole [\sorkin, \gperry]. In all 
the above cases one of the
directions of the D-2-brane, D-4-brane and D-6-brane is 
wrapped on the orbits of the $S^1$ isometry
of the KK-monopole.  This direction is transverse to 
the object (see for example [\erictwo]). 
 The six-dimensional \lq\lq gauge"
theory associated with the IIA KK-monopole has 
 excitations with sixteen degrees of freedom, one from
$(6_D|0_M)_A$, five from
$(2_D|0_M)_A$ and ten from $(4_D|0_M)_A$.

A BI type of action has been proposed for the IIA 
KK-monopole in [\erictwo] (we shall give more
details in the next section). It can be easily
 seen that the constant configuration
\constconf\  is also a solution of the IIA KK-monopole 
worldvolume theory. Therefore, this solution
has fifteen fluxes. The missing degree of freedom is 
due to the KK-monopole/D-6-brane bound state which does not
correspond to a solution of the effective worldvolume theory. 

In the IIB case, the bound states of D-branes with 
the IIB KK-monopole are T-duals
of the D-brane/NS-5-brane bound states of IIA theory 
along a direction transverse to the NS-5-brane
as 
$$
\eqalign{
(0_D|5_S)_A&{\buildrel  T\over\leftrightarrow} (1_D|0_M)_B
\cr
(2_D|5_S)_A&{\buildrel  T\over\leftrightarrow} (3_D|0_M)_B
\cr
(4_D|5_S)_A&{\buildrel  T\over\leftrightarrow} (5_D|0_M)_B\ .}
\eqn\twosixx
$$
In all the above cases one of the directions of
the D-1-brane, D-3-brane and D-5-brane is wrapped 
on the orbits of the $S^1$ isometry
of the KK-monopole.
The associated six-dimensional \lq\lq gauge" theory 
has  excitations with sixteen degrees
of freedom,
one is due to $(1_D|0_M)_B$, ten are due to 
$(3_D|0_M)_B$ and five are due to $(5_D|0_M)_B$. 
In this case, it is not straightforward to give the 
worldvolume description of the above bound
states since no BI type of action is known
 for the IIB KK-monopole. The
bulk counting suggests though that it 
should be similar to that of the IIA NS-5-brane.

\chapter{KK-monopole Worldvolume Solitons}

An effective worldvolume action for the M-theory KK-monopole has been
proposed in [\erictwo].  The construction of this action 
involves the use of  an isometry of the background
spacetime in such a way that the worldvolume theory
has the desirable number of 
degrees of freedom. To describe the bosonic part of
the worldvolume action for flat background, we 
first write the eleven-dimensional metric as
$$
ds^2(\bE^{(1,10)})=ds^2(\bE^{(1,9)})+dy^2\ ,
\eqn\kkone
$$
where $y$ is the coordinated adapted to the isometry 
and we choose the radius to be one.
Then the action in the static gauge  is
$$
I=\int d^7v \sqrt{|{\rm det}\big( \eta_{\alpha\beta}+\partial_\alpha X^r
\partial_\beta X^s \delta_{rs}+F_{\alpha\beta}\big)|}\ ,
\eqn\kkthree  
$$
where, $\{v^\alpha; \alpha=0,\dots,6\}$
 are the worldvolume coordinates,
$\eta$ is the seven-dimensional Minkowski metric, $F$ is the    
Born-Infeld field and  $\{X^r; r=1,2,3\}$ are the three 
transverse scalars which
are orthogonal to $y$.
The solitons
of this action are a 0-brane, a 3-brane and a 4-brane. 
The solutions are similar
to those of \zerobrane-\threebrane\ and we shall 
not give the expressions here. To find the
action of the IIA KK-monopole, we simply 
use a double reduction by 
identifying one of the worldvolume
directions, say
$v^6$, with one of the spacetime ones, say
$X^9$. This leads to the action
$$
I=\int d^6u \sqrt{|{\rm det}\big( \eta_{\mu\nu}+\partial_\mu
X^i\partial_\nu X^j\delta_{ij}+F_{\mu\nu}\big)|}\ ,
\eqn\kkfourr
$$
where $\{X^i; i=1, \dots,4\}=\{S,X^r; r=1,2,3\}$,
 $u^\mu=v^\mu$ for $\mu=0,\dots,5$ and $S$ is the
$v^6$ component of the BI potential.
This action is the same as that  of the  IIB
NS-5-brane in the static gauge. The brane solitons 
on the worldvolume of
the IIA KK-monopole are those of the IIB NS-5-brane.  
In particular, there is
a 0-brane soliton, a 2-brane soliton and a 
3-brane soliton. The expressions of these solutions
  are the same as those given in \zerobrane,
\twobrane\ and \threebrane.

From the bulk perspective, these solitons can be 
viewed as follows: (i) The 0-brane
soliton is associated with the intersection of a 
 D-2-brane with KK-monopole. The supergravity
 solution can be  found using the
T-duality chain 
$$
 (0|1_D, 5_S)_B{\buildrel  T\over\leftrightarrow} (0|2_D, 0_M)_A\ ,
\eqn\wmop
$$
where the T-duality direction $\tau$ 
is chosen  transverse to both
the D-string and NS-5-brane. The metric of the resulting
configuration in the string frame is
$$
\eqalign{
ds^2= -H^{-{1\over2}} dt^2+ H^{{1\over2}} 
ds^2(\bE^5)&
+ H^{-{1\over2}} V^{-1} (d\tau+\omega)^2
\cr &
+V H^{-{1\over2}} dx^2+ V H^{{1\over2}} dz d\bar z\ ,}
\eqn\newmwetricc
$$
where $H,V$ are the real parts of holomorphic functions 
$h, f$ in the coordinate $z$, respectively,
and $\omega={\rm Im} f dx$. This metric is the
 superposition of a D-2-brane 
and a  Gibbons-Hawking
metric [\hawkinga, \hawkingb] with the D-2-brane
 wrapped along the $\tau$ direction. (ii) The
2-brane  soliton is associated with the intetersection of a
D-4-brane with a KK-monopole. The 
supergravity solution can
 be found using the T-duality chain
$$
(2|3_D, 5_S)_B{\buildrel  T\over\leftrightarrow} (2|4_D, 0_M)_A\ ,
\eqn\kkfivee
$$
where the T-duality direction is chosen transverse to
both the D-3-brane and NS-5-brane. As in the previous case,
the solution is determined by two holomorphic functions.
(iii) The 3-brane soliton is the intersection 
of a IIA KK-monopole with a IIA NS-5-brane. 
The associated supergravity solution can
 be found using the T-duality chain
$$
(3|5_S,5_S)_B{\buildrel  T\over\leftrightarrow} (3|5_S, 0_M)_A\ ,
\eqn\kksixx
$$ 
where the  T-duality direction is chosen along 
a worldvolume direction of one of the NS-5-branes
transverse to the 3-brane intersection.  The solution is
again determined by two holomorphic functions.
 The 3-brane soliton cannot propagate in the
bulk since there is no 3-brane in IIA theory. 
All the above solutions preserve $1/4$ of bulk
supersymmetry.

As we have already mentioned in the previous section, 
the field equations of the worldvolume theory
of IIB KK-monopole are not known. However it is 
expected that the field equations will closely
resemble those of the IIA NS-5-brane. Therefore  
there should be a 3-brane and a 
string worldvolume solitons. From the bulk perspective, 
the 3-brane soliton can be realized as the
intersection of a NS-5-brane with a IIB KK-monopole.  
The supergravity solution can be constructed
by performing T-duality on the configuration of 
two IIA NS-5-branes intersecting on a 3-brane along
a worldvolume direction of one of the 5-branes 
transverse to the intersection, {\sl i.e.}
$$
(3|5_S,5_S)_A{\buildrel  T\over\leftrightarrow} (3|5_S,0_M)_B\ .
\eqn\kksevenn
$$
The string soliton can be realized as the intersection of  
D-3-brane with the IIB KK-monopole.  The supergravity
solution can be found using the T-duality chain
$$
(1|2_D,5_S)_A{\buildrel  T\over\leftrightarrow} (1_D|3_D,0_M)_B\ ,
\eqn\kkeightt
$$
where the T-duality direction is chosen transverse to both the
D-2-brane and IIA NS-5-brane.
The above two solutions
preserve $1/4$ of bulk supersymmetry and they are determined by
two holomorphic functions.

\chapter{T-duality and brane solitons}

The association of worldvolume solitons as 
brane boundaries or brane intersections allows the
investigation of their T-duality properties 
using the T-duality rules of IIA  and
IIB bulk theories. Therefore the T-duality 
transformations of the worldvolume brane solitons are
already contained in the bulk T-duality
 chains in sections three and five. However, as we shall see, it is
instructive to focus on the T-duality properties 
of the worldvolume solitons.
 Let
$p_{SA}$ and
$p_{KA}$ denote a p-brane worldvolume soliton  of 
IIA NS-5-brane and KK-monopole, 
respectively, and similarly
for IIB. 
Using the bulk T-duality chain
$$
(0|1_D,5_S)_B{\buildrel  T\over\leftrightarrow}
(1|2_D,5_S)_A {\buildrel  T\over\leftrightarrow}
(2|3_D,5_S)_B\ ,
\eqn\newwww
$$
 we find that
$$
0_{SB}{\buildrel  T\over\leftrightarrow} 
1_{SA}{\buildrel  T\over\rightarrow} 2_{SB}\ ,
\eqn\kkninee
$$
where the directions of the T-duality are 
along the worldvolume of the NS-5-brane.
It is clear from this that under T-duality 
the 0-brane, 1-brane and
2-brane worldvolume solitons transform like D-branes. The 
relation of these solitons under
T-duality also explains the similarity of the associated
 worldvolume solutions \zerobrane, \onebrane\ and \threebrane.
Similarly, using the bulk T-duality chain 
$$ 
(3|5_S,5_S)_B{\buildrel  T\over\leftrightarrow}(3|5_S,5_S)_A\ ,
\eqn\newnew
$$
 we
find that
$$
3_{SB}{\buildrel  T\over\leftrightarrow} 3_{SA}\ .
\eqn\kktenn
$$
As we have seen the worldvolume solutions of
$3_{SA}$ and $3_{SB}$ are the same.

Next using the bulk T-duality chains, 
$$
\eqalign{
(0|1_D,5_S)_B&{\buildrel  T\over\leftrightarrow}(0|0_D,0_M)_A
\cr
(2|3_D,5_S)_B&{\buildrel  T\over\leftrightarrow}(2|2_D,0_M)_A
\cr
(3|5_S,5_S)_B&{\buildrel  T\over\leftrightarrow}(3|5_S,0_M)_A\ ,}
\eqn\yeyeye
$$
  we find
$$
\eqalign{
0_{SB}&{\buildrel  T\over\leftrightarrow} 0_{KA}
\cr
2_{SB}&{\buildrel  T\over\leftrightarrow} 2_{KA}
\cr
3_{SB}&{\buildrel  T\over\leftrightarrow} 3_{KA}\ .}
\eqn\kkelevenn
$$
As we have already mentioned in the previous section
 the worldvolume solutions of the
relevant worldvolume soliton pairs are identical.

Finally, the T-duality properties of the worldvolume 
solitons of the IIB KK-monopole are
$$
\eqalign{
1_{SA}&{\buildrel  T\over\leftrightarrow} 1_{KB}
\cr
3_{SA}&{\buildrel  T\over\leftrightarrow} 3_{KB}\ .}
\eqn\kktwelvee
$$
The above T-duality transformations of the worldvolume solitons can be
combined by performing T-duality along different directions.
 
\chapter{Concluding Remarks}

We have shown that the fluxes
of the various six-dimensional \lq\lq gauge" 
theories are associated to bound states  below 
threshold of D-branes with the  NS-5-branes   and  
KK-monopoles. We then have used T-duality
to construct the associated supergravity solutions
 and we found that they
preserve $1/2$ of bulk supersymmetry.  In addition, we 
have investigated
the sectors of these theories that preserve 
$1/4$ of supersymmetry using the worldvolume
solitons of IIA and IIB NS-5-branes   and  
KK-monopoles. We have compared the
worldvolume solitons with  bulk configurations 
and found a new solution in IIA and IIB supergravity
with the interpretation of two NS-5-branes 
intersecting on a 3-brane. We have also presented some of
the worldvolume soliton solutions of M-theory 
and IIA KK-monopoles. In addition using the T-duality
rules of IIA and IIB theories, we have found 
some of the T-duality properties of the worldvolume
solitons and explained the similarities that 
some of the associated solutions have.

Most of our investigation was focused on the 
solitons that preserve $1/4$ of bulk supersymmetry. 
However many aspects of it can be 
easily extended to solitons that preserve
$1/8$ or less supersymmetry [\townb]. Our 
investigation can also be extended to the
non-abelian phase of the effective theories 
of the six-dimensional \lq\lq gauge" theories. In
particular, it is expected that there are
 soliton solution in this phase as well. For example, the
effective theory associated with the IIA 
KK-monopole and the effective theory associated with the
IIB NS-5-brane are expected to have instanton 
solutions which can be interpreted
as worldvolume string solitons. 

\vskip 1cm
\noindent{\bf Acknowledgments:}  I would like to thank G.W. Gibbons,
 M.B. Green and P.K. Townsend
for helpful discussions. I am supported by a University 
Research Fellowship from the Royal Society.
\vskip 1cm

\refout

\bye